\documentclass[epj,twocolumn]{webofc}
\usepackage[varg]{txfonts}   
\usepackage{graphicx}

\usepackage{booktabs} 
\usepackage{colortbl}

\usepackage{caption}
\usepackage{subcaption}

\wocname{epj}
\woctitle{Seismology of the Sun and the Distant Stars 2016}
\begin{document}
\title{Asteroseismic modelling of the Binary HD 176465}
%

\author{\firstname{B.} \lastname{Nsamba}\inst{1,2}\fnsep\thanks{\email{benard.nsamba@astro.up.pt}},
        \firstname{M. J. P. F. G.} \lastname{Monteiro}\inst{1,2},
         \firstname{T. L.} \lastname{Campante}\inst{3,5},
        \firstname{D. R.} \lastname{Reese}\inst{4,3},
        \firstname{T. R.} \lastname{White}\inst{5,6,7},
        \firstname{A.} \lastname{Garc\'{i}a Hern\'{a}ndez}\inst{1,2},\\
        \firstname{C.} \lastname{Jiang}\inst{1,2}
}

\institute{Instituto de Astrof\'{\i}sica e Ci\^{e}ncia do Espa\c{c}o, Universidade do Porto, Rua das
Estrelas, PT4150-762 Porto, Portugal
\and
Departamento de F\'{\i}sica e Astronomia, Faculdade de Ci\^{e}ncias da Universidade do Porto, Portugal 
\and
School of Physics and Astronomy, University of Birmingham, Birmingham B15 2TT, UK
\and
LESIA, Observatoire de Paris, PSL Research University, CNRS, Sorbonne Universités, UPMC Univ. Paris 06, Univ. Paris Diderot, Sorbonne Paris Cit\'{e}, 92195 Meudon, France
\and 
Stellar Astrophysics Centre, Department of Physics and Astronomy, Aarhus University, Ny Munkegade 120, DK-8000 Aarhus C, Denmark
\and
Institut f\"{u}r Astrophysik, Georg-August-Universit\"{a}t G\"{o}ttingen, Friedrich-Hund-Platz 1, 37077 G\"{o}ttingen, Germany
\and
Max-Planck-Institut f\"ur Sonnensystemforschung, Justus-von-Liebig-Weg 3, 37077 G\"ottingen, Germany
 }

\abstract{%
The detection and analysis of oscillations in binary star systems is critical in understanding stellar structure and evolution. This is partly because such systems
have the same initial chemical composition and age. Solar-like oscillations have been detected by Kepler in both components of the asteroseismic binary HD 176465. 
We present an independent modelling of each star in this binary system. Stellar models generated using MESA
(Modules for Experiments in Stellar Astrophysics) were fitted to both the observed individual frequencies and complementary spectroscopic parameters. 
The individual theoretical oscillation frequencies for the corresponding stellar models were obtained using GYRE as the pulsation code.
A Bayesian approach was applied to find the probability distribution functions of the stellar parameters using AIMS (Asteroseismic Inference on a Massive Scale)
as the optimisation code. The ages of HD 176465 A and HD 176465 B were found to be 2.81 $\pm$ 0.48 Gyr and 2.52 $\pm$ 0.80 Gyr, respectively.                     
These results are in agreement when compared to previous studies carried out using other asteroseismic modelling techniques and gyrochronology.
}
\maketitle
%
\section{Introduction}
\label{intro}
The accomplishments of the CoRoT and Kepler space telescopes have tremendously transformed asteroseismology of solar-like oscillations into an effective tool 
for characterising stars \cite{RefMI, RefCM}. Apart from the observation of solar-like oscillations in several hundreds of solar-type stars, a few 
binary systems with solar-like oscillations detected separately in both components have also been observed, e.g., 16 Cyg \cite{RefMT}, $\alpha$ Cen \cite{RefBd, RefKJ}.
Binary star systems play an important role in the understanding of stellar structure  and evolution. This is because such systems share the same formation
history, i.e., formed from the same molecular cloud therefore having the same initial chemical composition and approximately the same age.
\par
Different tools are currently employed in asteroseismic modelling of stars to derive fundamental stellar parameters \cite{RefSA}. 
The asteroseismic inference is achieved by matching model parameters to observed individual oscillation frequencies
or ratios of characteristic frequency separations, and spectroscopic parameters such as effective temperature and metallicity 
 \cite{RefMC}.
\par
The individual frequencies and spectroscopic parameters for the asteroseismic binary HD 176465 obtained with 
high-quality photometric observations using Kepler are presented in \cite{RefN}. In this article, we adopt this available data set to estimate a robust set of fundamental properties for
both components in this system by using the state-of-the-art AIMS\footnote{http://bison.ph.bham.ac.uk/spaceinn/aims/} optimisation tool based on a grid of
evolutionary models generated using MESA\footnote{http://mesa.sourceforge.net/}.
\par
This paper is organised as follows. Section \ref{sec-1} gives a brief description of our target and defines the observables to be
matched  with our generated models, while our model grid constituents and optimisation procedures are discussed in Section \ref{sec-2}. The 
results are discussed in Section \ref{sec-3} including a comparison with results from other asteroseismic fitting algorithms. Our 
conclusions are given in Section \ref{sec-4}.

\section{Target and data}
\label{sec-1}
HD  176465 is a nearby  binary system  displaying solar-like oscillations in both components. The Kepler space telescope observed the 
binary system in short-cadence mode (SC; 58.85s sampling), 
for 30 days during the asteroseismic survey phase (20-07-2009 to 19-08-2009, i.e., Quarter 2.2), and continuously from the 
end of the survey (37 months, Quarters 5 - 17). Refer to \cite{RefJJ} and \cite{RefGRA} for details on how the SC time series were prepared from raw 
observations and corrected to remove outliers and jumps.
Both components in the binary system have overlapping frequencies which have been 
extracted through detailed modelling of the power spectrum and analysis of the \'{e}chelle diagram (see \cite{RefN} for details).

\begin{table}[!ht]
\centering
\caption{Spectroscopic parameters of HD 176465 A and  HD 176465 B}   
\label{tab-spec}
\begin{tabular}{ | c | c | c |  }        
\hline\hline 
Parameter  & HD 176465 A  & HD 176465 B \\    
\hline\hline                       
[Fe/H](dex)               &  -0.30 $\pm$ 0.06       &  -0.30 $\pm$ 0.06   \\      
T$_{\mbox{eff}}$ (K)           & 5830 $\pm$ 90         &   5740 $\pm$ 90    \\      
\hline\hline                                  
\end{tabular}                   
\end{table}

In order to increase the precision of asteroseismic inference, we adopted both the  spectroscopic parameters (i.e., metallicity (Fe/H) 
and effective temperature (T$_{\mbox{eff}}$) in Table \ref{tab-spec}) 
and the seismic parameters (i.e., the observed individual frequencies in \cite{RefN}).   
A spectrum of HD 176465 was obtained from the ESPaDOnS spectrograph on the 
3.6-m Canada - France - Hawaii Telescope (See \cite{RefBrun} and \cite{RefMolend} for details on how the spectrum was analysed, and the fundamental parameters
and element abundances derived).

\section{Modelling}
\label{sec-2}
MESA version 7624 \cite{RefI, RefJ, RefK} was used to construct a grid of stellar models for 
each of the individual component in the binary system HD 176465. 
The stellar models used opacities from OPAL tables \cite{RefG} for temperatures above $10^4$ K. At lower temperatures,
tables from \cite{RefD}  were used. Nuclear reaction rates were obtained from tables provided by the NACRE collaboration \cite{RefA}. Specific
rates for $^{14}$N$(p,\gamma)^{15}$O were described by \cite{RefF} and for $^{12}$C$(\alpha,\gamma)^{16}$O by \cite{RefH}.
The standard grey Eddington atmosphere was used to integrate the atmospheric structure from the photosphere to an optical depth of
$\tau = 10^{-4}$. The mixing length theory \cite{RefC} was used to describe convection. No convection overshooting or semiconvection was included,
however diffusion was included to cater for gravitational settling of heavy elements according to the method of \cite{RefL}.
\par
The stellar grid contained a mass range M $\in$ [0.7, 1.15] in steps of 0.05 M$_\odot$, a mixing length parameter range $\alpha \in$ [1.5, 2.2] in steps of 0.1 and a
metal abundance range Z $\in$ [0.005, 0.021] in steps of 0.001. For every combination of M, Z, and $\alpha$, stellar evolution tracks were evolved by MESA from the 
pre-main sequence to the end of the main sequence. Over 60 models at different ages along each evolution track were generated. The oscillation
frequencies for each model were calculated using GYRE\footnote{https://bitbucket.org/rhdtownsend/gyre/} \cite{RefM}. The surface effects in the 
model frequencies were corrected using the two-term correction described by \cite{RefB} and implemented in AIMS.
\par
AIMS uses a Bayesian approach to find the probability distribution functions (PDFs) of the stellar parameters. By interpolating in the grid of models calculated by MESA,
AIMS generates a representative sample of models using Monte Carlo Markov Chain (MCMC) approach based on the 
Python package emcee.py\footnote{http://dan.iel.fm/emcee/current/} \cite{RefE}.
The results of the derived stellar parameters are presented in Figures ~\ref{mass} and ~\ref{radius}, and Tables ~\ref{tab-1} and ~\ref{tab-2}. 
\section{Discussion}
\label{sec-3}
Our results are compared with findings obtained using other modelling procedures and tools. These include the 

\begin{figure}[!ht]
  \centering
  \includegraphics[width=\hsize]{Benard_Nsamba_MassA.pdf}
\end{figure}
\vspace{-0.9cm}
\begin{figure}[iht]
  \centering
  \includegraphics[width=\hsize]{Benard_Nsamba_MassB.pdf}
\caption{ Comparison of our results with those obtained by several pipelines in \cite{RefN}. 
The red box corresponds to 1$\sigma$-error interval in our results.
}
\label{mass}
\end{figure}


\begin{figure*}[!ht]
\centering
\begin{minipage}{.5\textwidth}
  \centering
  \includegraphics[width=\hsize]{Benard_Nsamba_RadiusA.pdf}
\end{minipage}%
\begin{minipage}{.5\textwidth}
  \centering
  \includegraphics[width=\hsize]{Benard_Nsamba_RadiusB.pdf}
\end{minipage}
\caption{Comparison of our results with those obtained by several pipelines in \cite{RefN}. 
The red box corresponds to 1$\sigma$-error interval in our results.
}
\label{radius}
\end{figure*}


\begin{table*}
\begin{center}
\caption{Comparison of our results (AIMS) with those obtained using other modelling 
strategies (AMP, ASTFIT, BASTA, and MESA \cite{RefN}) for HD 176465 A}  
\label{tab-1}
\begin{tabular}{ | c | c | c | c | c | c | }        
\hline\hline 
Parameter  & AMP & ASTFIT  & BASTA & MESA& AIMS \\    
\hline\hline                       
 Mass (M$_\odot$)  &  0.930$\pm$ 0.04  & 0.952 $\pm$ 0.015      &  0.960$^{+0.010} _{-0.011}$   &  0.99 $\pm$ 0.02& 0.943 $\pm$ 0.011 {\footnotesize ($\pm$ 0.001)}  \\      
 Radius (R$_\odot$)     & 0.918 $\pm$ 0.015 & 0.927 $\pm$ 0.005      & 0.928$^{+0.006} _{-0.003}$  & 0.939 $\pm$ 0.006 &  0.924 $\pm$ 0.004 {\footnotesize ($\pm$ 0.0003)}\\
 Age (Gyr)        &  3.0 $\pm$ 0.4   & 3.2 $\pm$ 0.5       &   2.8 $\pm$ 0.3              & 3.01 $\pm$ 0.12      & 2.81 $\pm$ 0.48 {\footnotesize ($\pm$ 0.6)} \\
 Z                  &  0.0085 $\pm$ 0.0010 & 0.0103 $\pm$ 0.0010   & 0.011 $\pm$ 0.004       & 0.0094 $\pm$ 0.0009  & 0.012 $\pm$ 0.0010 {\footnotesize ($\pm$ 0.0004)}\\
 Y$_{i}$		   &   0.258 $\pm$ 0.024  &  0.262 $\pm$ 0.003    & 0.265 $\pm$ 0.002           & 0.23 $\pm$ 0.02        &    -----    \\
 $\alpha$            & 1.90 $\pm$ 0.18  & 1.80       &   1.791                              & 1.79        &  1.78 $\pm$ 0.11{\footnotesize ($\pm$ 0.01)}\\      
\hline\hline                                  
\end{tabular}
\end{center}
\vspace{-0.1cm}
{\small{Note: ----- The surface helium abundance (Y$_{i}$)} is not determined in AIMS}.
\end{table*}

\begin{table*}
\begin{center}
\caption{Comparison of our results (AIMS) with those obtained using other modelling 
strategies (AMP, ASTFIT, BASTA, and MESA \cite{RefN}) for HD 176465 B.} 
\label{tab-2}
\begin{tabular}{ | c | c | c | c | c | c | }        
\hline\hline 
 Parameter  & AMP & ASTFIT & BASTA & MESA & AIMS\\    
\hline\hline                        
Mass (M$_\odot$)  &  0.930$\pm$ 0.02 & 0.92 $\pm$ 0.02       &  0.929$^{+0.010} _{-0.011}$     &  0.97 $\pm$ 0.04    & 0.921 $\pm$ 0.015 {\footnotesize ($\pm$ 0.002)}\\      
Radius (R$_\odot$)     & 0.885 $\pm$ 0.006 & 0.883 $\pm$ 0.007      & 0.886$^{+0.003} _{-0.006}$    & 0.899 $\pm$ 0.013      &    0.883 $\pm$ 0.005 {\footnotesize ($\pm$ 0.001)}\\
Age (Gyr)        &  2.9 $\pm$ 0.5    & 3.4 $\pm$ 0.9             &   3.2 $\pm$ 0.4             & 3.18 $\pm$ 0.31       & 2.52 $\pm$ 0.80 {\footnotesize ($\pm$ 0.1)}\\   
Z                  &  0.0085 $\pm$ 0.0007 & 0.0096 $\pm$ 0.0011   & 0.011 $\pm$ 0.004           & 0.0122 $\pm$ 0.0011    & 0.011 $\pm$ 0.0012 {\footnotesize ($\pm$ 0.0002)}\\
Y$_i$		   &   0.246 $\pm$ 0.013  &  0.262 $\pm$ 0.003    & 0.265 $\pm$ 0.002           & 0.24 $\pm$ 0.04        &    -----    \\
$\alpha$            & 1.94 $\pm$ 0.12    & 1.80                   &   1.791                     & 1.79         & 1.79 $\pm$ 0.11 {\footnotesize ($\pm$ 0.01)}\\      
\hline\hline                                   
\end{tabular}
\end{center}
\vspace{-0.1cm}
{\small{Note: ----- The surface helium abundance (Y$_{i}$)} is not determined in AIMS}.
\end{table*}


\noindent
Asteroseismic Modelling Portal (AMP), the Aarhus STellar Evolution Code {\small{(ASTEC)}} Fitting method (ASTFIT), the Bayesian Stellar Algorithm (BASTA), and  MESA code.
Refer to \cite{RefN} and  references therein for details on the different tools.
MESA can also be used for optimisation purposes because it contains a module capable of fitting a set of asteroseismic 
constraints (results in Tables \ref{tab-1} and \ref{tab-2}, under column MESA). 
In this study, MESA was only used to generate a grid of stellar models as described in Section \ref{sec-2}.
The input physics used varies across the set of modelling tools used. Furthermore, each modelling technique uses different
stellar evolution codes, pulsation codes and/or optimization techniques. 
The results in Tables \ref{tab-1} and \ref{tab-2} include only statistical uncertainties returned by each pipeline. 
\par
The error bars in brackets under column AIMS (Tables \ref{tab-1} and \ref{tab-2}) were obtained using an unreduced ~$\chi^2$ value in the likelihood function.
These errors are consistent from a statistical point of view, i.e., they reflect how the observational errors propagate to the derived stellar parameters.
The main error bars used were obtained using the overall reduced ~$\chi^2 $ which is inconsistent from
a statistical point of view. This option, however, generates error bars comparable to those obtained using other optimization tools. 
\par
The different modelling tools converge on models for HD 176465 A with a mean mass of 0.95 $\pm$ 0.02 M$_\odot$ and a mean radius of 0.93 $\pm$ 0.01 R$_\odot$.
For HD 176465 B, they converge with a mean mass of 0.93 $\pm$ 0.02 M$_\odot$ and a mean radius of 0.89 $\pm$ 0.01  R$_\odot$.
Figures \ref{mass} and \ref{radius} show that results from MESA are slightly higher regarding the stellar masses and radii compared to results from other modelling tools. 
The initial helium abundance of the best-fitting MESA models are lower than the primordial value from standard Big Bang nucleosynthesis (Y = 0.2482 $\pm$ 0.0007)\cite{RefSteig}.
This may explain why masses and radii of MESA models are larger compared to other modeling tools.
\par
The age of HD 176465 A and HD 176465 B was found to be
2.81 $\pm$ 0.48 Gyr and 2.52 $\pm$ 0.80 Gyr  respectively. This is consistent within 1$\sigma$-error interval. 
All models agree on the ages of both stars within the uncertainties even when no restrictions were made on the ages during the different modelling procedures.
This is expected assuming both stars were formed at approximately the same time and from the same molecular cloud.
The gyrochronology relations  based on \cite{RefGarcia} calibrations were used in \cite{RefN} and the age of the binary was found to be 3.2 $^{+1.2} _{-0.8}$ Gyr. 
This is in good agreement with results obtained using the asteroseismic modelling tools.
\par
The metal abundances for HD 176465 A and HD 176465 B are 0.012 $\pm$ 0.0010 and 0.011 $\pm$ 0.0012, respectively.
This is consistent within 1$\sigma$-error interval. This is expected assuming that both stars were formed from
the same molecular cloud with approximately the same chemical composition.

\section{Conclusions}
\label{sec-4}
We derived precise fundamental stellar parameters using state of the art modelling tools for the asteroseismic binary system. 
The obtained results were compared with those available from other modeling tools \cite{RefN}. The methods agree on the 
stellar properties in almost all cases.
\par
Binary star systems provide important tests of stellar structure and evolution. They can be used to provide important
constraints in the asteroseismic modelling techniques especially when orbital parameters are known. 
For HD 176465, position measurements were made in the Washington Double Star
Catalog \cite{RefMAS}. Unfortunately the fraction of the orbit covered is insufficient to constrain the total mass \cite{RefN}.
\par
Only a handful of main sequence binary systems have been made available from 
space missions such as CoRoT and Kepler. Future missions such as TESS \cite{RefTLC} and PLATO \cite{RefTx} are expected to increase 
asteroseismic data for similar systems. This will help in setting tight constraints on the physics used in stellar evolution
and asteroseismology.
\\
\\
\noindent
{
\small
{\bf Acknowledgements}: We thank the MESA community for the engaging conversations about the MESA project. B. Nsamba acknowledges support
from Funda\c{c}\~{a}o para a Ci\^{e}ncia e a Tecnologia (FCT, Portugal) under the Grant ID: PD/BD/113744/2015.
TLC acknowledges the support of the UK Science and Technology Facilities Council (STFC).
Furthermore, A. Garc\'{i}a Hern\'{a}ndez acknowledges support from Funda\c{c}\~{a}o para a Ci\^{e}ncia e a Tecnologia (FCT, Portugal)
through the fellowship SFRH/BPD/80619/2011 from the EC Project SPACEINN (FP7-SPACE-2012-312844). 
GYRE has been developed with support from the Astronomy and Astrophysics Grants (AAG) program of 
the National Science Foundation, through awards AST-0908688 and AST-0904607.
The ``Asteroseismic Inference on a Massive Scale'' (AIMS) project was developed at the University of Birmingham by Daniel R. Reese as 
one of the deliverables for the SPACEINN network. The SPACEINN network is funded by the European Community’s Seventh Framework Programme (FP7/2007-2013) 
under grant agreement no. 312844.
}

%
%
\bibliographystyle{plain}
\bibliography{bibliography.bib}

\end{document}